\documentclass[conference]{IEEEtran}
\IEEEoverridecommandlockouts
\usepackage{lipsum}
 \newcommand{\diff}{\mathrm{d}}

\newcommand{\R}{\mathbb{R}}

\newcommand{\E}{\mathbb{E}}
\usepackage{amsthm} 
\theoremstyle{definition}

\usepackage{array}
\usepackage{makecell}

\usepackage{blindtext}
\usepackage{xcolor}
\usepackage{cite}
\usepackage{amsmath,amssymb,amsfonts}
\usepackage{algorithmic}
\usepackage{graphicx}
\usepackage{textcomp}
\usepackage{xcolor,soul}
\def\BibTeX{{\rm B\kern-.05em{\sc i\kern-.025em b}\kern-.08em
    T\kern-.1667em\lower.7ex\hbox{E}\kern-.125emX}}

\begin{document}


\title{A Unified Framework for Calculating First Arrival Position Density in Molecular Communication\\

}

\author{\IEEEauthorblockN{Yen-Chi Lee}
\IEEEauthorblockA{\textit{Hon Hai (Foxconn) Research Institute} \\
Taipei, Taiwan \\
yen-chi.lee@foxconn.com}
\and
\IEEEauthorblockN{ 
Jen-Ming Wu
}
\IEEEauthorblockA{\textit{Hon Hai (Foxconn) Research Institute} \\
Taipei, Taiwan \\
jen-ming.wu@foxconn.com}
\and
\IEEEauthorblockN{Min-Hsiu Hsieh}
\IEEEauthorblockA{\textit{Hon Hai (Foxconn) Research Institute} \\
Taipei, Taiwan \\
min-hsiu.hsieh@foxconn.com}
}

\maketitle

\begin{abstract}

\color{black}
This paper introduces a novel unified framework for calculating the first arrival position (FAP) density in diffusion-based molecular communication (MC) systems with fully-absorbing receivers, applicable to any spatial dimension or receiver shape. We validate the effectiveness of our proposed method by presenting two concrete examples from the MC literature: the 2D line receiver and 3D spherical receiver. Our approach successfully reproduces existing results on FAP density and further enhances the original formula for the 2D line receiver by accommodating arbitrary drift directions. The proposed framework offers a comprehensive and versatile solution for determining FAP density in diffusive MC channels with fully-absorbing receivers, paving the way for more robust and efficient system designs.
\color{black}

\end{abstract}

\begin{IEEEkeywords}
molecular communication (MC), absorbing receiver, first arrival position (FAP), stochastic differential equation (SDE), It\^{o} diffusion.
\end{IEEEkeywords}

\section{Introduction} 
\label{section:intro}


The challenge of transmitting information over long distances while maintaining the fidelity of the information has been present throughout human history, from ancient times to the present day. While modern communication systems have addressed this problem using electromagnetic (EM) signals, such techniques encounter limitations in nano-scale applications due to restrictions in wavelength, antenna size, and energy requirements \cite{nakano2013molecular, guo2015molecular}.

\color{black}
To overcome the limitations of electromagnetic signals, molecular signals have been proposed as an alternative for nanonetwork applications. In molecular communication (MC) systems, information is conveyed through tiny molecules, referred to as message molecules (MM), that act as information carriers \cite{akyildiz2008nanonetworks,yeh2012new}. To transport these MMs through the physical channel toward the receiver, a propagation mechanism is necessary. Various forms of propagation mechanisms exist, such as diffusion-based \cite{berg2018random}, flow-based \cite{kadloor2012molecular}, or an engineered transport system like molecular motors \cite{moore2006design,gregori2010new}. Of these, \textit{diffusion-based} propagation (sometimes in combination with advection and chemical reaction networks) has been the most commonly used model in MC systems and will be the main focus of this study.
It should be noted that in this paper, we use the terms ``molecules'' and ``particles'' interchangeably to refer to the MMs, as their shape is not relevant to our discussion. 
\color{black}

In diffusion-based MC systems, information can be transmitted by modulating various physical properties of the MMs \cite{farsad2016comprehensive,kuran2011modulation,kuran2020survey,hsieh2013asynchronous}. Upon reaching the vicinity of the receiver, signaling molecules can be observed and processed to extract the necessary information for detection and decoding \cite{jamali2019channel}. The reception mechanism of an MC receiver can be classified into two categories: i) \text{passive reception}, and ii) \text{active reception}. The simplest active reception model is the \textit{fully-absorbing receiver} \cite{yilmaz2014three}, which can measure the hitting time and position of each MM \cite{akdeniz2018molecular}, and subsequently remove the MM upon receipt \cite{lee2016distribution}. In this study, we will focus on (fully) absorbing receivers.

Developing accurate channel models is a key challenge in the field of MC \cite{jamali2019channel}. In MC systems that employ an absorbing receiver, accurately characterizing particle arrival position or time distribution is crucial for effective system design.
There are three types of first-arrival models in MC: time-modulation, position-modulation, and joint position-time modulation. Time-modulation encodes information in the releasing time of a MM, and the probability density of the ``random delay'' due to molecule propagation mechanism is used to characterize this channel \cite{srinivas2012molecular, yilmaz2014three, li2014capacity}. Position-modulation encodes information in the emission position of a MM, and the probability density of the ``random position deviation'' is used to model this channel \cite{lee2016distribution, pandey2018molecular}. Joint position-time modulation considers both position and time, and is used to solve inter-symbol interference issues \cite{akdeniz2018molecular}. Various first-arrival densities have been derived for different receiver shapes and dimensions.

This study focuses on characterizing first-arrival-position (FAP) channels. Previous works such as \cite{redner2001guide} have utilized a stochastic process known as the ``first-passage process" to obtain the FAP distribution for 3D spherical receivers. Alternatively, \cite{pandey2018molecular} studied the FAP distribution for 2D line receivers using the image method, similar to electric potential theory \cite{jackson1999classical}. In contrast, we propose a novel unified approach to solve the FAP density finding problem applicable for various receiver shapes and dimensions, deviating from the approaches proposed in \cite{redner2001guide} and \cite{pandey2018molecular}. Specifically, we derive the concise relation:
\begin{equation}
f_{Y|X}(\mathbf{y}|\mathbf{x}) = \left| \dfrac{\partial G}{\partial \mathbf{n}_\mathbf{y}}(\mathbf{x},\mathbf{y}) \right|,
\end{equation}
and the new method (with comparisons to the old method) is summarized in Table \ref{table:compare-gpt}.
\color{black}
In Section IV, we demonstrate the effectiveness of our method by solving the FAP density finding problem for two specific cases: the 2D line receiver and 3D spherical receiver. These two examples show that our new approach can readily reproduce these previously reported findings. Notably, for the 2D line receiver, we not only confirm the original formula \cite[Eq.~(19)]{pandey2018molecular} but also enhance the existing result by enabling arbitrary drift directions.
\color{black}

The paper is organized as follows: Section~\ref{sec:model} presents the system model and channel model utilized in this study. Section~\ref{section:generator} introduces our novel unified approach for finding FAP density. Section~\ref{sec:case} presents two concrete channels from MC literature as examples, and demonstrates that our approach can readily reproduce the existing results about FAP density. Finally, Section~\ref{sec:conclude} concludes the paper.
\section{System Model and Channel Model}\label{sec:model}
\subsection{System Model}

\begin{figure}[htbp]
\centerline{\includegraphics[width=0.5\textwidth]{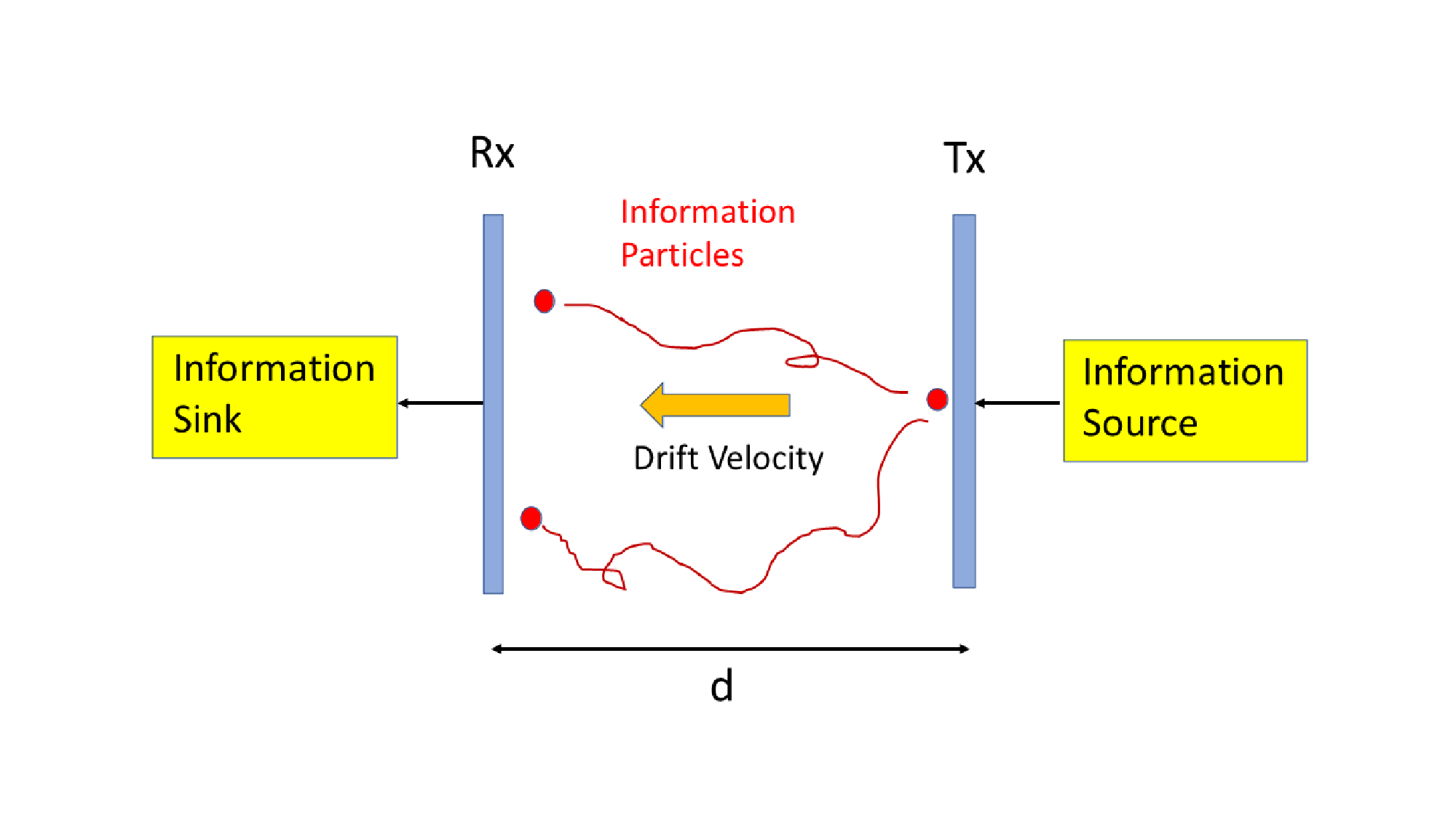}}
\caption{
This figure illustrates a 2D FAP channel with line-shape Tx and Rx, where the transmitter line is assumed to be transparent, allowing particles to move through it without experiencing any force. The emission point is located on the transmitter line.
}
\label{fig:2Da}
\end{figure}

\begin{figure}[htbp]
\centerline{\includegraphics[width=0.5\textwidth]{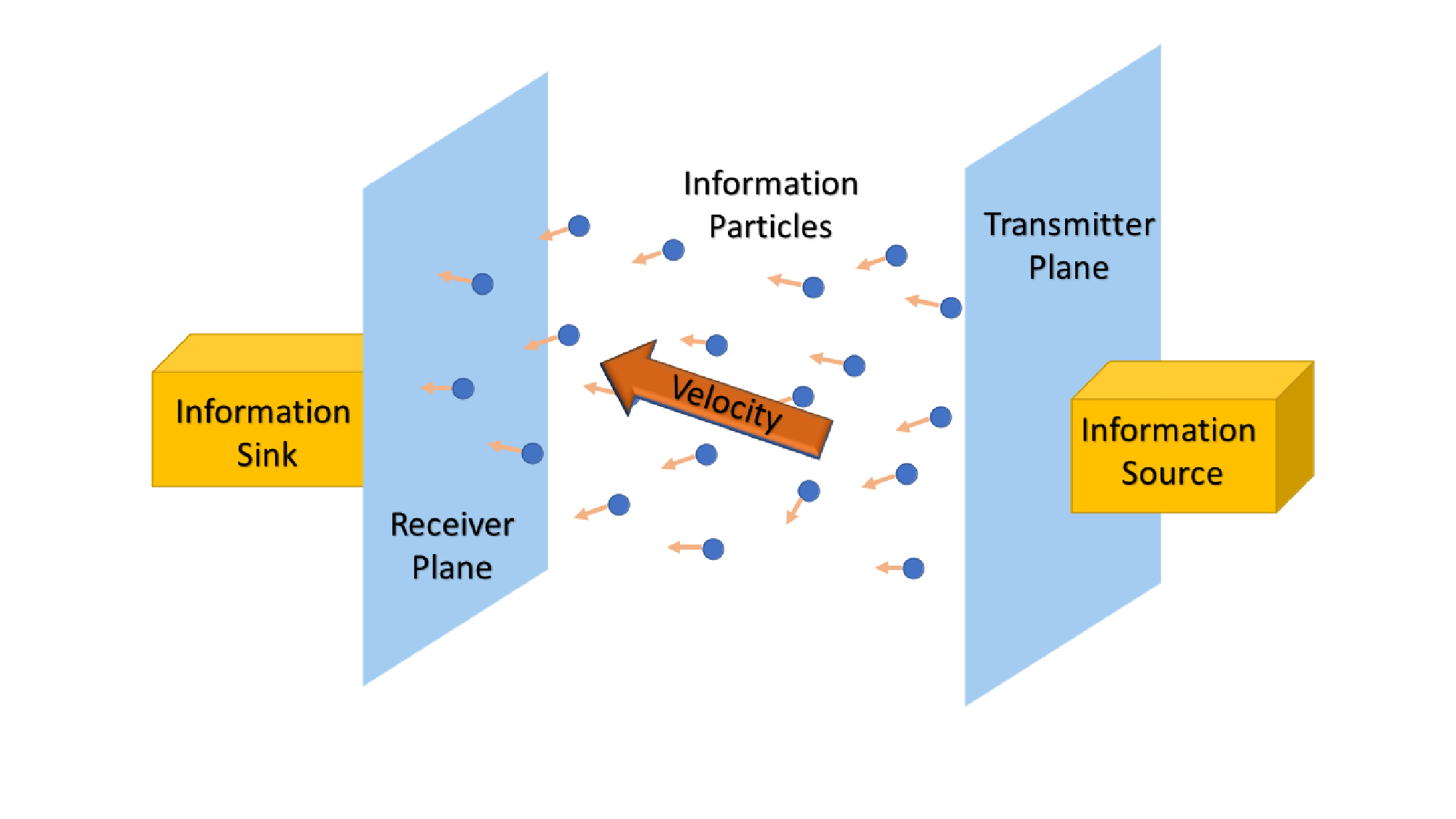}}
\caption{
This figure illustrates a 3D FAP channel with plane-shape Tx and Rx, where the transmitter plane is assumed to be transparent, allowing particles to move through it without experiencing any force. The emission point is located on the transmitter plane.
}
\label{fig:3Da}
\end{figure}

\color{black}
We have chosen a system model that is general enough to encompass the majority of MC channels related to FAP densities. Our approach considers a MC system situated in an $n$-dimensional space, comprising of a transmitter (Tx) and an absorbing receiver (Rx), with the background medium being fluid. The Tx emits a message molecule containing position information, while the Rx records its first interaction point before removing it from the medium. The molecules propagate via diffusion after being released. The distance between the emission point and the center of the receiver is denoted by $d$. We impose the condition that particles cannot go inside the Rx. The domain in which particles can travel is denoted by $\Omega$, and the fully-absorbing boundary of the Rx is referred to as $\partial \Omega$. We provide two examples of such MC systems in 2D and 3D spaces, as shown in Figure~\ref{fig:2Da} and Figure~\ref{fig:3Da}, respectively.
\color{black}

\subsection{Channel Model}

\color{black}
In the literature on MC simulations, there are two distinct perspectives for modeling diffusion channels. The macroscopic viewpoint utilizes the diffusion equation, also known as the heat equation, to depict the evolution of the concentration distribution of message molecules. This approach falls under the category of partial differential equation (PDE) models \cite{keener2021biology}. In contrast, the microscopic viewpoint models the kinematics of a message molecule using a random process described by It\^{o} diffusion \cite{oksendal2013stochastic,calin2015informal}.
For a more comprehensive understanding of these two viewpoints, readers can refer to classical potential theory, as detailed in \cite{doob1984classical}, or consult insightful review papers on MC, such as \cite{farsad2016comprehensive} or \cite{jamali2019channel}. In the following paragraphs, we will provide a brief description of each viewpoint.
\color{black}

\color{black}
From a macroscopic perspective, the diffusion channel can be described by the following diffusion-advection equation \eqref{ad-eq} in the presence of a background drift velocity. To perform MC system analysis, it is essential to obtain the concentration field $c(\mathbf{r},t)$ of message molecules at a specific spatial location $\mathbf{r}$ and time $t$. According to Fick's law of diffusion \cite{jamali2019channel}, the temporal evolution of $c(\mathbf{r},t)$ can be described as:
\begin{equation}
\partial_{t} c\left(\mathbf{r},t;\mathbf{r}_{0}\right)+\mathbf{v}(\mathbf{r}, t) \cdot \nabla c\left(\mathbf{r},t;\mathbf{r}_{0}\right)=D \Delta c\left(\mathbf{r},t;\mathbf{r}_{0}\right),
\label{ad-eq}
\end{equation}
where $\mathbf{r}_{0}$ denotes the starting point of diffusion, $\mathbf{v}$ represents the background velocity field, $\nabla$ and $\Delta:=\nabla^{2}$ are the gradient and Laplacian operators, respectively, and $D$ is the diffusion coefficient. The value of $D$ is dependent on factors such as temperature, fluid viscosity, and the molecule's Stokes radius, as outlined in \cite{farsad2016comprehensive}. Note that in physics, the term ``field" often refers to a function of space and time.
\color{black}

\color{black}
From a microscopic perspective, the suitable model for an individual trajectory $X_t$ of a message molecule is an It\^{o} diffusion process. In the following discussion, we will assume that the parameters in the diffusion channel vary slowly over time, allowing our molecular channel to be viewed as an approximately time-invariant system \cite{jamali2019channel}. This assumption corresponds to the time-homogeneous It\^{o} diffusion model.

A time-homogeneous It\^{o} diffusion process is a stochastic process that satisfies a stochastic differential equation (SDE) of the form:
\begin{equation}
\diff X_t = b(X_t)\diff t + \sigma(X_t) \diff B_t,
\label{eq:IP}
\end{equation}
where $B_t$ is an $n$-dimensional standard\footnote{Here, ``standard" refers to zero-mean and unit-variance.} Brownian motion. The first term, $b(X_t)\diff t$, is the deterministic component, which describes the drift effect induced by an external potential field. The second term, $\sigma(X_t) \diff B_t$, represents the random fluctuations caused by the constant bombardment of background molecules.
\color{black}

\section{Main Result: A Novel Approach for Efficiently Determining FAP Density} \label{section:generator}

In this section, we present a novel and unified approach for determining the FAP density. To achieve this goal efficiently, we introduce several powerful stochastic analytical tools, such as the generator of diffusion semigroup and Dynkin's formula.

\color{black}
Our new approach for finding the FAP density can be summarized in three steps:

\begin{enumerate}
\item Firstly, we remove the time-dependency from the FAP density by utilizing the generator of the diffusion semigroup that corresponds to the fluid media between Tx and Rx. This is because the FAP density is closely related to the steady-state of the diffusion phenomenon.

\item Secondly, we solve a boundary value problem (BVP) by combining the elliptic-type differential operator with appropriate boundary conditions, which are determined by the receiver mechanics. Representation formulas for commonly encountered boundary conditions can be found in PDE handbooks such as \cite{polyanin2001handbook}. Since there are many well-known results for solutions of elliptic PDEs, even in high-dimensional spaces (see \cite{grohs2022deep}), in many cases, it is not necessary to derive representation formulas from scratch.

\item Finally, we apply the Dynkin's formula to obtain the FAP density directly from the Green's function $G(\textbf{y},\textbf{x} )$. This avoids the complicated integration process with respect to the time variable $t$ which is required in most other related FAP works (e.g., \cite{pandey2018molecular}). Specifically, the FAP density can be computed as $f_{Y|X}(\textbf{y}|\textbf{x})=
\left| \frac{\partial G(\textbf{y},\textbf{x} )}{\partial \mathbf{n}_{\textbf{y}}} \right|$ once the elliptic Green's function $G(\textbf{y},\textbf{x})$ is known.
\end{enumerate}
\color{black}

\color{black}
A comparison between the old and new methods for determining the FAP density is shown in Table \ref{table:compare-gpt}. The ``old method" refers to those methods that solve heat equations directly, and an example can be found in \cite{pandey2018molecular}. In the old method, a parabolic-type PDE is used, which is time-dependent, while the new method removes the time-dependency by considering the generator of the diffusion semigroup. Additionally, the old method often requires time-integration after calculating the flux density, while the new method directly obtains the FAP density using formula \eqref{eq:main}.
\color{black}

\begin{table*}[t]
\caption{Comparison Between Old and New Method of Calculating FAP Density}
\centering
\begin{tabular}{|c|p{5.5cm}|p{8cm}|}
\hline
& Old Method
& New Method \\ \hline
Step 1 & Calculates the free space Green's function for parabolic-type PDE. & Removes time-dependency via considering the generator of the diffusion semigroup. \\ \hline
Step 2 & Accommodates the free space solution to absorbing boundary conditions (e.g., using the image method). & Solves elliptic-type BVP with appropriate boundary conditions. \\ \hline
Step 3 & Calculates flux density and does time-integration. & Obtains FAP density directly from the Green's function $G(\textbf{y},\textbf{x} )$. No time integration is required. \\ \hline
\end{tabular}
\label{table:compare-gpt}
\end{table*}

\color{black}
The objective of the remainder of this section is to validate the effectiveness of our proposed FAP density finding approach. To achieve this, we have divided the verification process into two distinct parts, which are presented in separate subsections, as follows:
\color{black}

\subsection{Infinitestimal Generator of It\^{o} Diffusion}

\color{black}
In physics, an It\^{o} diffusion is commonly utilized to describe the Brownian motion of a particle in a fluid medium under the influence of a potential. In mathematics, an $n$-dimensional It\^{o} diffusion
is a stochastic process that satisfies a specific type of SDE, given by
$
\diff X_t = b(X_t)\diff t + \sigma(X_t) \diff B_t,
$
as shown in \eqref{eq:IP}.
We assume that $b(\cdot)$ and $\sigma(\cdot)$ are constants, allowing the corresponding MC system to be regarded as a time-invariant system.
\color{black}

\color{black}
For each It\^{o} diffusion $X_t$, a corresponding operator called the infinitesimal generator, or simply generator, can be associated. Let $\mathcal{D}(A)$ denote the domain of the generator $A$. We define the notation $\E^{\mathbf{x}}[\cdot]$ as taking expectation conditioned on $X_{0}=\mathbf{x}$, so that
$
\mathbb{E}^{\mathbf{x}}[f(X_t)] := \mathbb{E}[f(X_t)|X_0=\mathbf{x}].
$
The generator $A$ (which operates on function $f$) of a process $X_t$ can be defined as
\begin{equation}
Af=A\{f(\mathbf{x})\}:=\lim\limits_{t\searrow 0} \dfrac{\mathbb{E}^{\mathbf{x}}[f(X_t)]-f(\mathbf{x})}{t}
\text{\ for any\ } f\in \mathcal{D}(A).
\label{eq:def-generator}
\end{equation}
For time-homogeneous It\^{o} processes $X_t$, the time evolution is a Markov process.
By letting
$
T_t := \E^{\mathbf{x}}[f(X_t)],
$
we can obtain a semigroup of operators $\mathcal{T}=\{T_t\}_{t\geq 0}$.
Thinking from another perspective, the term $Af = \lim\limits_{t\searrow 0} \dfrac{T_t f - f}{t}$ can be regarded as the infinitesimal linear incremental term of $f$ through the semigroup evolution.
\color{black}

\color{black}
To explicitly calculate the generator $A$ based on \eqref{eq:IP},
we assume that $f$ is of class $C^2$, which means that it has both a continuous first derivative and a continuous second derivative. Using Taylor expansion and It\^{o}'s formula\footnote{It\^{o}'s formula and Taylor expansion of an It\^{o} process are standard results in stochastic analysis. For more details, see \cite{oksendal2013stochastic,calin2015informal}.}, we obtain:
\begin{align}
\begin{split}
\diff f\left(X_{t}\right)=\ &f^{\prime}\left(X_{t}\right) \diff X_{t}+\frac{1}{2} f^{\prime \prime}\left(X_{t}\right) \diff \langle X\rangle_{t}\\
=\ &\left[b\left(X_{t}\right) f^{\prime}\left(X_{t}\right)+\frac{\sigma^{2}\left(X_{t}\right)}{2} f^{\prime \prime}\left(X_{t}\right)\right] \diff t\\
&+f^{\prime}\left(X_{t}\right) \sigma\left(X_{t}\right) \diff B_{t},
\end{split}
\label{eq:te_modified}
\end{align}
where $\diff \langle X \rangle_t$ represents the quadratic variation of the random process $X_t$, see \cite{calin2015informal}. By plugging \eqref{eq:te_modified} into \eqref{eq:def-generator} and using the fact that $\mathbb{E}[\diff B_t]=0$, we obtain an explicit expression for the generator $A$ as:
\begin{equation}
A\{f(\mathbf{x})\}=b(\mathbf{x})f'(\mathbf{x})+\dfrac{\sigma^2(\mathbf{x})}{2}f''(\mathbf{x}),
\label{eq:ItoGe_modified}
\end{equation}
where $\mathbf{x}$ can be viewed as the starting point of a diffusion process, and $\mathbf{x}\in \Omega$. From \eqref{eq:ItoGe_modified}, we can see that $A$ is a second-order differential operator of elliptic type.
\color{black}


\color{black}
We proceed to consider a BVP with an unknown function $u$, given by:
\begin{equation}
\left\{\begin{array}{ll}
Au=0 & \text { in } \Omega \\
u=g & \text { on } \partial \Omega
\end{array}\right..
\label{eq:BVP1-1}
\end{equation}
Here, $A$ denotes a partial differential operator and $g$ is the prescribed boundary data, which is also a $C^2$ function. For elliptic-type partial differential operators, representation formulas for commonly encountered boundary conditions can be found in PDE handbooks such as \cite{polyanin2001handbook}. Since many well-known results for solutions of elliptic PDEs exist, even in high-dimensional spaces (see \cite{grohs2022deep}), it is often unnecessary to derive representation formulas from scratch.
\color{black}


\subsection{A Novel Relation Between FAP Density and Elliptic Green's Function}\label{subsection:D}

\color{black}
The equation that defines an infinitesimal generator is recalled as \eqref{eq:def-generator}. By rearranging the terms in \eqref{eq:def-generator}, we can obtain the following integral expression:
\begin{equation}\label{generator-int}
\E^{\mathbf{x}}\left[f\left(X_{t}\right)\right]=f(\mathbf{x})+\E^{\mathbf{x}}\left[\int_{0}^{t} A \{f\left(X_{s}\right)\} \diff s\right]
.\end{equation}
It is important to note that the variable $t$ appearing in \eqref{generator-int} is deterministic, and not a random variable.
\color{black}

\color{black}
In the field of stochastic analysis, Dynkin's formula is a theorem that provides valuable information about a diffusion process at a given stopping time $\tau$. Specifically, suppose $f$ is a $C^2$ function and let $\tau$ be a stopping time such that $\E^{\mathbf{x}}[\tau]<+\infty$. Then, Dynkin's formula can be expressed as:
\begin{equation}
\E^{\mathbf{x}}\left[f\left(X_{\tau}\right)\right]=f(\mathbf{x})+\E^{\mathbf{x}}\left[\int_{0}^{\tau} A \{f\left(X_{s}\right)\} \diff s\right].
\label{formula-dynkin}
\end{equation}
Notably, the variable $\tau$ is now a random variable. Additional details regarding this formula can be found in \cite{calin2015informal}.
\color{black}

\color{black}
Let $g$ denote a smooth function defined on the boundary $\partial \Omega$. Using the Chapman-Kolmogorov equation, we can express the expected value of $g(X_\tau)$ as follows:
\begin{equation}
\E^{\mathbf{x}}[g(X_\tau)]=\E[g(X_\tau)\mid X_0=\mathbf{x}]
=\int_{\partial \Omega} f_{Y|X}(\mathbf{y}| \mathbf{x})g(\mathbf{y})\diff \mathbf{y}.
\label{eq:22}
\end{equation}
Here, $\mathbf{x}\in \Omega$ represents the starting point of the diffusion, while the hitting position $\mathbf{y}$ belongs to the receiving boundary, i.e., $\mathbf{y}\in\partial \Omega$. It is worth noting that the conditional probability density function $f_{Y|X}$ in \eqref{eq:22} is exactly the desired FAP density on the receiver boundary $\partial \Omega$. Our remaining task is to determine $f_{Y|X}$ using Green's function.
\color{black}

\color{black}
Suppose we have a solution $u(\mathbf{x})$ for the BVP in \eqref{eq:BVP1-1} with prescribed boundary data $g$ such that $Au=0$ inside $\Omega$. By setting $f(\mathbf{x})=u(\mathbf{x})$ in \eqref{formula-dynkin}, we obtain:
$
\int_0^\tau A\{u(X_s)\} \diff s =0.
$
Here, $0<s<\tau$ and $\tau$ represents the first hitting time. The physical interpretation of this condition is that $X_s$ is located inside $\Omega$ prior to the hitting event, which leads to $A\{u(X_s)\}=0$. Furthermore, since $u(\mathbf{x})$ coincides with $g(\mathbf{x})$ on the boundary, we can express the left-hand side of \eqref{formula-dynkin} as $\E^{\mathbf{x}}[g(X_\tau)]$. Combining these two facts yields:
\begin{equation}
\E^{\mathbf{x}}[g(X_\tau)]=u(\mathbf{x})+0=
\int_{\partial \Omega} f_{Y|X}(\mathbf{y}| \mathbf{x})g(\mathbf{y})\diff \mathbf{y}.
\label{main-1}
\end{equation}
This holds true for any $\mathbf{x}\in \Omega$.
\color{black}


\color{black}
For elliptic-type BVPs defined in a domain $V$ with boundary $\partial V$, it is generally true that \cite{polyanin2001handbook}:
\begin{equation}
u(\mathbf{x})=\int_{V} \Phi(\mathbf{y}) G(\mathbf{x}, \mathbf{y}) \diff V_{\mathbf{y}}+\int_{\partial V} g(\mathbf{y}) H(\mathbf{x}, \mathbf{y}) \diff S_{\mathbf{y}}.
\label{general-green}
\end{equation}
Here, $\Phi$ represents the source term, $d V_{\mathbf{y}}$ is the volume element, $d S_{\mathbf{y}}$ is the surface element, and $H(\mathbf{x}, \mathbf{y})$ depends on the type of boundary conditions under consideration. For MC systems without molecule reproduction or annihilation inside the channel, we can set $\Phi(\mathbf{y})=0$. For deriving FAP density, the Dirichlet-type boundary conditions are most appropriate, which correspond to:
$
H(\mathbf{x},\mathbf{y})=-\dfrac{\partial G}{\partial \mathbf{n}_\mathbf{y}}(\mathbf{x},\mathbf{y})
$.
This yields the simplified form of \eqref{general-green}:
\begin{equation}
u(\mathbf{x})=\int_{\partial V}
\dfrac{\partial G}{\partial \mathbf{n}_\mathbf{y}}(\mathbf{x},\mathbf{y})
g(\mathbf{y}) \diff S{\mathbf{y}}.
\label{general-green-2}
\end{equation}
Finally, by comparing \eqref{main-1} and \eqref{general-green-2}, we can establish a fundamental relation between the FAP density and the elliptic-type Green's function:
\begin{equation}
f_{Y|X}(\mathbf{y}|\mathbf{x}) = \left| \dfrac{\partial G}{\partial \mathbf{n}_\mathbf{y}}(\mathbf{x},\mathbf{y}) \right|
,
\label{eq:main}
\end{equation}
which represents the main result of this study.
\color{black}

\section{Calculating FAP Density in Specific Scenarios}\label{sec:case}
This section outlines the application of formula \eqref{eq:main} and the novel methodology presented in Section~\ref{section:generator} to compute the FAP density in practical scenarios that are relevant to MC. Notably, in subsection~\ref{subsec:2D-line}, we not only replicate the findings in \cite{pandey2018molecular}, but also extend the drift velocity to accommodate any direction. Consequently, formula \eqref{2Dresult} constitutes a new result in the MC domain.

\subsection{Absorbing 2D Line Receiver}\label{subsec:2D-line}

\color{black}
The study in \cite{pandey2018molecular} examines a MC channel where the fluid medium experiences constant drift from the transmitter towards the receiver, specifically in the longitudinal direction. The absorbing receiver, as illustrated in Figure~\ref{fig:2Da}, is assumed to have a linear shape. Utilizing our novel approach, we aim to determine the FAP density on the line receiver in this subsection. Our derivations not only recover the formula presented in \cite{pandey2018molecular}, but also accommodate drift in any direction.
\color{black}

\color{black}
To solve the 2D FAP density problem, we consider an It\^{o} diffusion $X_t\in\R^2$ with a semigroup generator
\begin{equation}
A := \sum_{i=1}^2 v_i \dfrac{\partial}{\partial x_i} + \dfrac{\sigma^2}{2} \sum_{i=1}^2 \dfrac{\partial^2}{\partial x_i^2},
\end{equation}
where $v_i$ denotes the $i$-th component of the drift velocity $\mathbf{v}$. We denote the Laplacian operator in 2D by $\Delta_2$ and consider the following BVP in Cartesian coordinates:
\begin{equation}
\left\{\begin{array}{ll}
Au=\sum_{i=1}^{2} v_{i} \frac{\partial u}{\partial x_{i}}+\frac{\sigma^2}{2} \Delta_2 u=0 & \text { in } \Omega \\
u=g & \text { on } \partial \Omega
\end{array}\right.
\label{BVP-2D},
\end{equation}
where $u$ is a (dummy) unknown function, $\Omega$ is the domain of \eqref{BVP-2D} defined by $\Omega=\mathbb{R}^{2} \cap\left\{x_{2}>0\right\}$, and $\partial\Omega$ is the boundary defined by $\partial\Omega=\mathbb{R}^{2} \cap\left\{x_{2}=0\right\}$. Here, we use the notation $x_j$ to indicate the $j$-th component of the position vector $\mathbf{x}=(x_1,\cdots,x_n)$ in an $n$-dimensional space. We temporarily set $\sigma^2=1$.
\color{black}

\color{black}
To simplify the problem further, we introduce a change of variable $w(\mathbf{x})=\gamma(\mathbf{x}) u(\mathbf{x})$ and utilize an alternative BVP described by the Helmholtz operator, where
$
\gamma(\mathbf{x}):=\exp {\mathbf{v} \cdot \mathbf{x}}=\exp \left(v_{1} x_{1}+v_{2} x_{2}\right),
$
yielding:
\begin{equation}
\left\{\begin{array}{ll}
H(w)=\Delta_2 w-s^{2} w=0 & \text { in } \Omega \\
w=\tilde{g} & \text { on } \partial \Omega
\end{array}\right.
,
\label{BVP-2D2}
\end{equation}
where
$
s=|\mathbf{v}|
=\sqrt{v_1^2+v_2^2}
.$
According to \cite{polyanin2001handbook}, the solution to equation \eqref{BVP-2D2} is given by:
\begin{equation}
w(x_1, x_2) =\int_{-\infty}^{\infty} \int_{-\infty}^{\infty} f(\xi)\left[\frac{\partial}{\partial \eta} G(x_1, x_2, \xi, \eta)\right]_{\eta=0} \diff \xi.
\label{eq:handbook2}
\end{equation}
In the above equation, the Green's function has the form
$
\begin{array}{c}
G(x_1, x_2, \xi, \eta)=
\frac{1}{2\pi}[K_0(s\rho_1)-K_0(s\rho_2)]
\end{array}
$, where $\rho_1:=\sqrt{(x_1-\xi)^{2}+(x_2-\eta)^{2}}$ and $\rho_2:=\sqrt{(x_2-\xi)^{2}+(x_2+\eta)^{2}}$.
Note that $K_0(\cdot)$ denotes the modified Bessel function of the second kind, of order $0$.
\color{black}

\color{black}
Next, we set $\mathbf{x}=(x_1,d)$ and $\mathbf{y}=(\xi,0)$. Utilizing the methodology presented in Section~\ref{section:generator}, we can express $w(\mathbf{x})$ as follows:
\begin{equation}
w(\mathbf{x})=\int_{\partial \Omega}\left|\frac{\partial G(\mathbf{x}, \mathbf{y})}{\partial \mathbf{n}_\mathbf{y}}\right| \tilde{g}(\mathbf{y}) \diff \mathbf{y}.
\label{new-repre2D}
\end{equation}
In \eqref{new-repre2D}, the integral kernel can be explicitly written down through direct calculations:
\begin{align}
\begin{split}
&\left|\frac{\partial G(\mathbf{x}, \mathbf{y})}{\partial \mathbf{n}_\mathbf{y}}\right|_{\mathbf{x}=(x_1,d),\ \mathbf{y}=(\xi,0)}
\\
=&\ \exp \left\{v_{1} \xi-v_1 x_1 - v_2 d \right\}
\left[\dfrac{\partial G}{\partial \rho_1}\dfrac{\partial \rho_1}{\partial \eta}+\dfrac{\partial G}{\partial \rho_2}\dfrac{\partial \rho_2}{\partial \eta}\right]_{\eta=0}\\
=&\ \dfrac{|v|d}{\pi}\exp\{-v_2 d\} \exp\{-v_1(x_1-\xi)\}\\
&\ \cdot\dfrac{K_1\left(|\mathbf{v}|\sqrt{(x_1-\xi)^2+d^2}\right)}{\sqrt{(x_1-\xi)^2+d^2}}.
\end{split}
\end{align}
In this context, $K_1(\cdot)$ denotes the modified Bessel function of the second kind, of order 1, and $d$ represents the distance from the point of emission to the line of reception, as illustrated in Figure~\ref{fig:2Da}.
For the general case $\sigma^2\neq 1$, we only need to replace $v_i$ with $\frac{v_i}{\sigma^2}$, which gives the desired FAP density that allows arbitrary drift directions:
\begin{align}
\begin{split}
f_{Y \mid X}(\mathbf{y}|\mathbf{x})=&\ \dfrac{|\mathbf{v}|d}{\sigma^2\pi}\exp\left\{\dfrac{-v_2 d}{\sigma^2}\right\} \exp\left\{\dfrac{-v_1(x_1-\xi)}{\sigma^2}\right\}\\ &\cdot\dfrac{K_1\left(\dfrac{|\mathbf{v}|}{\sigma^2}\sqrt{(x_1-\xi)^2+d^2}\right)}{\sqrt{(x_1-\xi)^2+d^2}},
\end{split}
\label{2Dresult}
\end{align}
where $\mathbf{x}=(x_1,d)$, $\mathbf{y}=(\xi,0)$.
\color{black}

\color{black}
To recover the restrictive version presented in \cite{pandey2018molecular}, we set $v_1=0$ in \eqref{2Dresult}, which yields:
\begin{align}
\begin{split}
&f_{Y \mid X}(\mathbf{y}|\mathbf{x})\\
=&\ 
\dfrac{|\mathbf{v}|d}{\sigma^2\pi}
\exp\left\{\dfrac{-v_2 d}{\sigma^2}\right\} \dfrac{K_1\left(\dfrac{|\mathbf{v}|}{\sigma^2}\sqrt{(x_1-\xi)^2+d^2}\right)}{\sqrt{(x_1-\xi)^2+d^2}}\\
=&\ 
\dfrac{|\mathbf{v}|d}{2\pi D}
\exp\left\{\dfrac{-v_2 d}{2D}\right\} \dfrac{K_1\left(\dfrac{|\mathbf{v}|}{2D}\sqrt{(x_1-\xi)^2+d^2}\right)}{\sqrt{(x_1-\xi)^2+d^2}}
,
\end{split}
\label{eq:2DFAP-corrected}
\end{align}
where we use the relation $\sigma^2 = 2D$. Therefore, our result \eqref{2Dresult} improves the formula \cite[Eq.(19)]{pandey2018molecular} by allowing arbitrary drift directions.
\color{black}


\subsection{Absorbing 3D Spherical Receiver}

\color{black}
In this subsection, we consider the 3D spherical receiver case with the assumption of zero drift velocity, which is the same setting as used in \cite{akdeniz2018molecular}. Since the result we are going to show is already known in MC, we present a very brief version of how to derive it using our methods, and suppress most of the details. The main goal of this subsection is to demonstrate the capability of our method to handle receivers with various shapes.
\color{black}

\color{black}
For the 3D FAP density problem, we consider an It\^{o} diffusion $X_t\in\R^3$ with zero drift, so all $v_i=0$. The corresponding generator $A$ is simply the 3D Laplacian. We can express the 3D Laplace equation in the spherical coordinate system as
\begin{align}
\begin{split}
\frac{1}{r^{2}} \frac{\partial}{\partial r}\left(r^{2} \frac{\partial w}{\partial r}\right)
&+\frac{1}{r^{2} \sin \theta} \frac{\partial}{\partial \theta}\left(\sin \theta \frac{\partial w}{\partial \theta}\right)\\
&+\frac{1}{r^{2} \sin ^{2} \theta} \frac{\partial^{2} w}{\partial \phi^{2}}=0
\end{split},
\end{align}
where $r=\sqrt{x^2+y^2+z^2}$. Assuming a prescribed boundary data $g(\theta,\phi)$ at the surface ($r=R$) of the spherical receiver with radius $R$, i.e.,
\begin{equation}
    w=g(\theta,\varphi) \text{\ at\ } r=R,
\end{equation}
the solution of the outer problem for $r\geq R$ can be obtained as \cite{polyanin2001handbook}:
\begin{align}
\begin{split}
w(r, \theta, \phi)=&\frac{R}{4 \pi} \int_{0}^{2 \pi} \int_{0}^{\pi} g\left(\theta_{0}, \phi_{0}\right)\cdot\\ &\frac{r^{2}-R^{2}}{\left(r^{2}-2 rR c(\theta,\phi;\theta_0,\phi_0)+R^{2}\right)^{3 / 2}} \sin \theta_{0} \diff \theta_{0} \diff \phi_{0},
\end{split}
\label{eq:43}
\end{align}
where
$
c(\theta,\phi;\theta_0,\phi_0) := \cos\theta \cos\theta_0 + \sin\theta \sin\theta_0 \cos(\phi-\phi_0).
$ 
\color{black}


\color{black}
To conclude this subsection, the comparison between formula \eqref{eq:43} and equation \eqref{new-repre2D} allows us to determine
$\left| \dfrac{\partial G}{\partial \mathbf{n}_\mathbf{y}}(\mathbf{x},\mathbf{y}) \right|$, which is exactly the FAP density $f_{Y|X}(\mathbf{y}|\mathbf{x})$ due to our proposed formula \eqref{eq:main}. That is, we have already obtained the marginal angular distribution of the hitting position of molecules, which was expressed in \cite[Eq.(3)]{akdeniz2018molecular} and \cite[Eq.(6.3.3a)]{redner2001guide}. This demonstrates the versatility of our methodology in handling various receiver shapes, including spherical receivers.
\color{black}
\section{Conclusion}\label{sec:conclude}

\color{black}
In this paper, we have presented a comprehensive solution to the first-arrival-position (FAP) density problem in molecular communication for fully-absorbing receivers. Our approach establishes a connection between macroscopic partial differential equation (PDE) models and microscopic stochastic differential equation (SDE) models.

We have mathematically verified our method in Section~\ref{section:generator} using stochastic analysis tools, such as the generator of a semigroup and Dynkin's formula. This verification led to a concise expression, represented by \eqref{eq:main}, which relates the FAP density to the elliptic-type Green's function. Notably, this expression is applicable to any spatial dimension and receiver shape.

Additionally, we have demonstrated the effectiveness of our approach by providing concrete examples from the molecular communication literature. Specifically, using the 2D line receiver and 3D spherical receiver, we have shown that our approach can reproduce known results on FAP density. Moreover, for the 2D line receiver, our approach improves upon the existing formula by accommodating arbitrary drift directions.

In conclusion, our method offers a powerful tool for solving the FAP density problem in diffusive molecular communication channels, particularly for fully-absorbing receivers, and has broad applicability to various spatial dimensions and receiver shapes.
\color{black}

\bibliographystyle{./IEEEtran}
\bibliography{main}

\begin{thebibliography}{10}
\providecommand{\url}[1]{#1}
\csname url@samestyle\endcsname
\providecommand{\newblock}{\relax}
\providecommand{\bibinfo}[2]{#2}
\providecommand{\BIBentrySTDinterwordspacing}{\spaceskip=0pt\relax}
\providecommand{\BIBentryALTinterwordstretchfactor}{4}
\providecommand{\BIBentryALTinterwordspacing}{\spaceskip=\fontdimen2\font plus
\BIBentryALTinterwordstretchfactor\fontdimen3\font minus
  \fontdimen4\font\relax}
\providecommand{\BIBforeignlanguage}[2]{{%
\expandafter\ifx\csname l@#1\endcsname\relax
\typeout{** WARNING: IEEEtran.bst: No hyphenation pattern has been}%
\typeout{** loaded for the language `#1'. Using the pattern for}%
\typeout{** the default language instead.}%
\else
\language=\csname l@#1\endcsname
\fi
#2}}
\providecommand{\BIBdecl}{\relax}
\BIBdecl

\bibitem{nakano2013molecular}
T.~Nakano, A.~W. Eckford, and T.~Haraguchi, \emph{Molecular
  communication}.\hskip 1em plus 0.5em minus 0.4em\relax Cambridge University
  Press, 2013.

\bibitem{guo2015molecular}
W.~Guo, C.~Mias, N.~Farsad, and J.-L. Wu, ``Molecular versus electromagnetic
  wave propagation loss in macro-scale environments,'' \emph{IEEE Transactions
  on Molecular, Biological and Multi-Scale Communications}, vol.~1, no.~1, pp.
  18--25, 2015.

\bibitem{akyildiz2008nanonetworks}
I.~F. Akyildiz, F.~Brunetti, and C.~Bl{\'a}zquez, ``Nanonetworks: A new
  communication paradigm,'' \emph{Computer Networks}, vol.~52, no.~12, pp.
  2260--2279, 2008.

\bibitem{yeh2012new}
P.-C. Yeh, K.-C. Chen, Y.-C. Lee, L.-S. Meng, P.-J. Shih, P.-Y. Ko, W.-A. Lin,
  and C.-H. Lee, ``A new frontier of wireless communication theory:
  diffusion-based molecular communications,'' \emph{IEEE Wireless
  Communications}, vol.~19, no.~5, pp. 28--35, 2012.

\bibitem{berg2018random}
H.~C. Berg, \emph{Random walks in biology}.\hskip 1em plus 0.5em minus
  0.4em\relax Princeton University Press, 2018.

\bibitem{kadloor2012molecular}
S.~Kadloor, R.~S. Adve, and A.~W. Eckford, ``Molecular communication using
  brownian motion with drift,'' \emph{IEEE Transactions on NanoBioscience},
  vol.~11, no.~2, pp. 89--99, 2012.

\bibitem{moore2006design}
M.~Moore, A.~Enomoto, T.~Nakano, R.~Egashira, T.~Suda, A.~Kayasuga, H.~Kojima,
  H.~Sakakibara, and K.~Oiwa, ``A design of a molecular communication system
  for nanomachines using molecular motors,'' in \emph{Fourth Annual IEEE
  International Conference on Pervasive Computing and Communications Workshops
  (PERCOMW'06)}.\hskip 1em plus 0.5em minus 0.4em\relax IEEE, 2006, pp. 6--pp.

\bibitem{gregori2010new}
M.~Gregori and I.~F. Akyildiz, ``A new nanonetwork architecture using
  flagellated bacteria and catalytic nanomotors,'' \emph{IEEE Journal on
  selected areas in communications}, vol.~28, no.~4, pp. 612--619, 2010.

\bibitem{farsad2016comprehensive}
N.~Farsad, H.~B. Yilmaz, A.~Eckford, C.-B. Chae, and W.~Guo, ``A comprehensive
  survey of recent advancements in molecular communication,'' \emph{IEEE
  Communications Surveys \& Tutorials}, vol.~18, no.~3, pp. 1887--1919, 2016.

\bibitem{kuran2011modulation}
M.~{\c{S}}. Kuran, H.~B. Yilmaz, T.~Tugcu, and I.~F. Akyildiz, ``Modulation
  techniques for communication via diffusion in nanonetworks,'' in \emph{2011
  IEEE international conference on communications (ICC)}.\hskip 1em plus 0.5em
  minus 0.4em\relax IEEE, 2011, pp. 1--5.

\bibitem{kuran2020survey}
M.~{\c{S}}. Kuran, H.~B. Yilmaz, I.~Demirkol, N.~Farsad, and A.~Goldsmith, ``A
  survey on modulation techniques in molecular communication via diffusion,''
  \emph{IEEE Communications Surveys \& Tutorials}, vol.~23, no.~1, pp. 7--28,
  2020.

\bibitem{hsieh2013asynchronous}
Y.-P. Hsieh, Y.-C. Lee, P.-J. Shih, P.-C. Yeh, and K.-C. Chen, ``On the
  asynchronous information embedding for event-driven systems in molecular
  communications,'' \emph{Nano Communication Networks}, vol.~4, no.~1, pp.
  2--13, 2013.

\bibitem{jamali2019channel}
V.~Jamali, A.~Ahmadzadeh, W.~Wicke, A.~Noel, and R.~Schober, ``Channel modeling
  for diffusive molecular communication—a tutorial review,''
  \emph{Proceedings of the IEEE}, vol. 107, no.~7, pp. 1256--1301, 2019.

\bibitem{yilmaz2014three}
H.~B. Yilmaz, A.~C. Heren, T.~Tugcu, and C.-B. Chae, ``Three-dimensional
  channel characteristics for molecular communications with an absorbing
  receiver,'' \emph{IEEE Communications Letters}, vol.~18, no.~6, pp. 929--932,
  2014.

\bibitem{akdeniz2018molecular}
B.~C. Akdeniz, N.~A. Turgut, H.~B. Yilmaz, C.-B. Chae, T.~Tugcu, and A.~E.
  Pusane, ``Molecular signal modeling of a partially counting absorbing
  spherical receiver,'' \emph{IEEE Transactions on Communications}, vol.~66,
  no.~12, pp. 6237--6246, 2018.

\bibitem{lee2016distribution}
Y.-C. Lee, C.-C. Chen, P.-C. Yeh, and C.-H. Lee, ``Distribution of first
  arrival position in molecular communication,'' in \emph{2016 IEEE
  International Symposium on Information Theory (ISIT)}.\hskip 1em plus 0.5em
  minus 0.4em\relax IEEE, 2016, pp. 1033--1037.

\bibitem{srinivas2012molecular}
K.~V. Srinivas, A.~W. Eckford, and R.~S. Adve, ``Molecular communication in
  fluid media: The additive inverse gaussian noise channel,'' \emph{IEEE
  transactions on information theory}, vol.~58, no.~7, pp. 4678--4692, 2012.

\bibitem{li2014capacity}
H.~Li, S.~M. Moser, and D.~Guo, ``Capacity of the memoryless additive inverse
  gaussian noise channel,'' \emph{IEEE Journal on Selected Areas in
  Communications}, vol.~32, no.~12, pp. 2315--2329, 2014.

\bibitem{pandey2018molecular}
N.~Pandey, R.~K. Mallik, and B.~Lall, ``Molecular communication: The first
  arrival position channel,'' \emph{IEEE Wireless Communications Letters},
  vol.~8, no.~2, pp. 508--511, 2018.

\bibitem{redner2001guide}
S.~Redner, \emph{A guide to first-passage processes}.\hskip 1em plus 0.5em
  minus 0.4em\relax Cambridge university press, 2001.

\bibitem{jackson1999classical}
J.~D. Jackson, ``Classical electrodynamics,'' 1999.

\bibitem{keener2021biology}
J.~P. Keener, \emph{Biology in time and space: a partial differential equation
  modeling approach}.\hskip 1em plus 0.5em minus 0.4em\relax American
  Mathematical Soc., 2021, vol.~50.

\bibitem{oksendal2013stochastic}
B.~{\O}ksendal, \emph{Stochastic differential equations: an introduction with
  applications}.\hskip 1em plus 0.5em minus 0.4em\relax Springer Science \&
  Business Media, 2013.

\bibitem{calin2015informal}
O.~Calin, \emph{An informal introduction to stochastic calculus with
  applications}.\hskip 1em plus 0.5em minus 0.4em\relax World Scientific, 2015.

\bibitem{doob1984classical}
J.~L. Doob and J.~Doob, \emph{Classical potential theory and its probabilistic
  counterpart}.\hskip 1em plus 0.5em minus 0.4em\relax Springer, 1984, vol.
  549.

\bibitem{polyanin2001handbook}
A.~D. Polyanin, \emph{Handbook of linear partial differential equations for
  engineers and scientists}.\hskip 1em plus 0.5em minus 0.4em\relax Chapman and
  hall/crc, 2001.

\bibitem{grohs2022deep}
P.~Grohs and L.~Herrmann, ``Deep neural network approximation for
  high-dimensional elliptic pdes with boundary conditions,'' \emph{IMA Journal
  of Numerical Analysis}, vol.~42, no.~3, pp. 2055--2082, 2022.

\end{thebibliography}


\end{document}